# Interplay of Fluorescence and Phosphorescence in Organic Biluminescent Emitters

*Caterin Salas Redondo,[1,2] Paul Kleine,[1] Karla Roszeitis,[1] Tim Achenbach,[1] Martin Kroll,[1] Michael Thomschke,[1] and Sebastian Reineke*[1,2].*

[1] Dresden Integrated Center for Applied Physics and Photonic Materials (IAPP) and Institute for Applied Physics, Technische Universität Dresden, Nöthnitzer Straße 61, D-01187, Germany.

[2] Center for Advancing Electronics Dresden (cfaed), Technische Universität Dresden, Würzburger Straße 46, D-01187 Dresden, Germany.

AUTHOR INFORMATION

**Corresponding Author**

*Sebastian Reineke. E-mail: reineke@iapp.de

# ABSTRACT

Biluminescent organic emitters show simultaneous fluorescence and phosphorescence at room temperature. So far, the optimization of the room temperature phosphorescence (RTP) in these materials has drawn the attention of research. However, the continuous wave operation of these emitters will consequently turn them into systems with vastly imbalanced singlet and triplet populations, which is due to the respective excited state lifetimes. This study reports on the exciton dynamics of the biluminophore NPB (N,N'-di(1-naphthyl)-N,N'-diphenyl-(1,1-biphenyl)-4,4-diamine). In the extreme case, the singlet and triplet exciton lifetimes stretch from 3 ns to 300 ms, respectively. Through sample engineering and oxygen quenching experiments, the triplet exciton density can be controlled over several orders of magnitude allowing to studying exciton interactions between singlet and triplet manifolds. The results show, that singlet-triplet annihilation reduces the overall biluminescence efficiency already at moderate excitation levels. Additionally, the presented system represents an illustrative role model to study excitonic effects in organic materials.

TEXT

The electronic properties of organic molecules are characterized by distinct spin manifolds as a consequence of joint effects of highly localized excitations and, compared to inorganic materials, low dielectric constants [1]. The results are typical energetic splitting of first singlet and triplet excited states, respectively, in the order of a few hundred meV [1]. Organic molecules typically only show efficient fluorescence, i.e. the emission from the singlet state, as it is an allowed transition [2]. On the contrary, the transition strength from the triplet state to ground state under emission of a photon (phosphorescence) is extremely weak, because it is quantum mechanically forbidden requiring the participating electron to undergo a spin flip [2]. In the vast of organic molecules, a low rate of radiative transition in the triplet manifold remains due to weak coupling, however it is outcompeted by non-radiative recombination at room temperature. As a consequence, the triplet state in organic molecules is typically considered a 'dark' state. Still, phosphorescence of organic molecules can be achieved by two means (cf. Figure 1b): (i) enhancement of the radiative phosphorescence rate $k_{r,P}$ to outcompete the non-radiative channel $k_{nr,P}$ or (ii) suppress $k_{nr,P}$ in favor of a dominating $k_{r,P}$. The first route is followed in a concerted way for the development of phosphorescent emitters for organic light-emitting diodes (OLEDs), owing the fact that about 75% of the formed excitons are born as triplets [3]. Here, heavy metal atoms are incorporated in the core of a molecular structure to enhance $k_{r,P}$ due to strong spin-orbit coupling (SOC) [4]. A weaker version of the same strategy is the increase of $k_{r,P}$ making use of SOC induced by a metal atom in the vicinity of the emitter (external heavy atom effect) [5]. Additionally, special molecular designs can equally be in favor of a moderately enhanced phosphorescence rate [6-8]. Route (ii) is most effectively realized by incorporating the emitter molecules at low concentration in rigid polymer hosts [9] or alternative matrices [10]. Such room temperature phosphorescence (RTP) from purely organic molecules has attracted much attention recently [11-16], because the persistent nature of the phosphorescence is a unique luminescent feature allowing for various novel applications [17-18]. In all studies to date, the sole focus has been put on the enhancement of the phosphorescence of organic molecules. However, light is emitted in such systems from both their singlet (fluorescence) and triplet excited states (Figure 1b and c), which is due to a competition of $k_{ISC}$, $k_{r,F}$ and $k_{nr,F}$ (i.e. intersystem crossing, fluorescence radiative, and non-radiative rates). Principally, because the oscillator strength is much larger in the singlet manifold, excitation of the molecules is almost exclusively realized via absorption to singlet states ($S_0 \to S_1$). This biluminescence, or dual-state emission, can be efficient for any given mixing ratio between fluorescence and phosphorescence, if both $k_{r,F} \gg k_{nr,F}$ and $k_{r,P} \gg k_{nr,P}$. Essentially, the typical lifetimes of fluorescence (nanoseconds) and phosphorescence (milliseconds to seconds) differ greatly with direct influence on the overall population of the molecules. Of course, these novel emitters are of particular interest for potential applications, because their luminescence characteristics can be retained at room temperature.

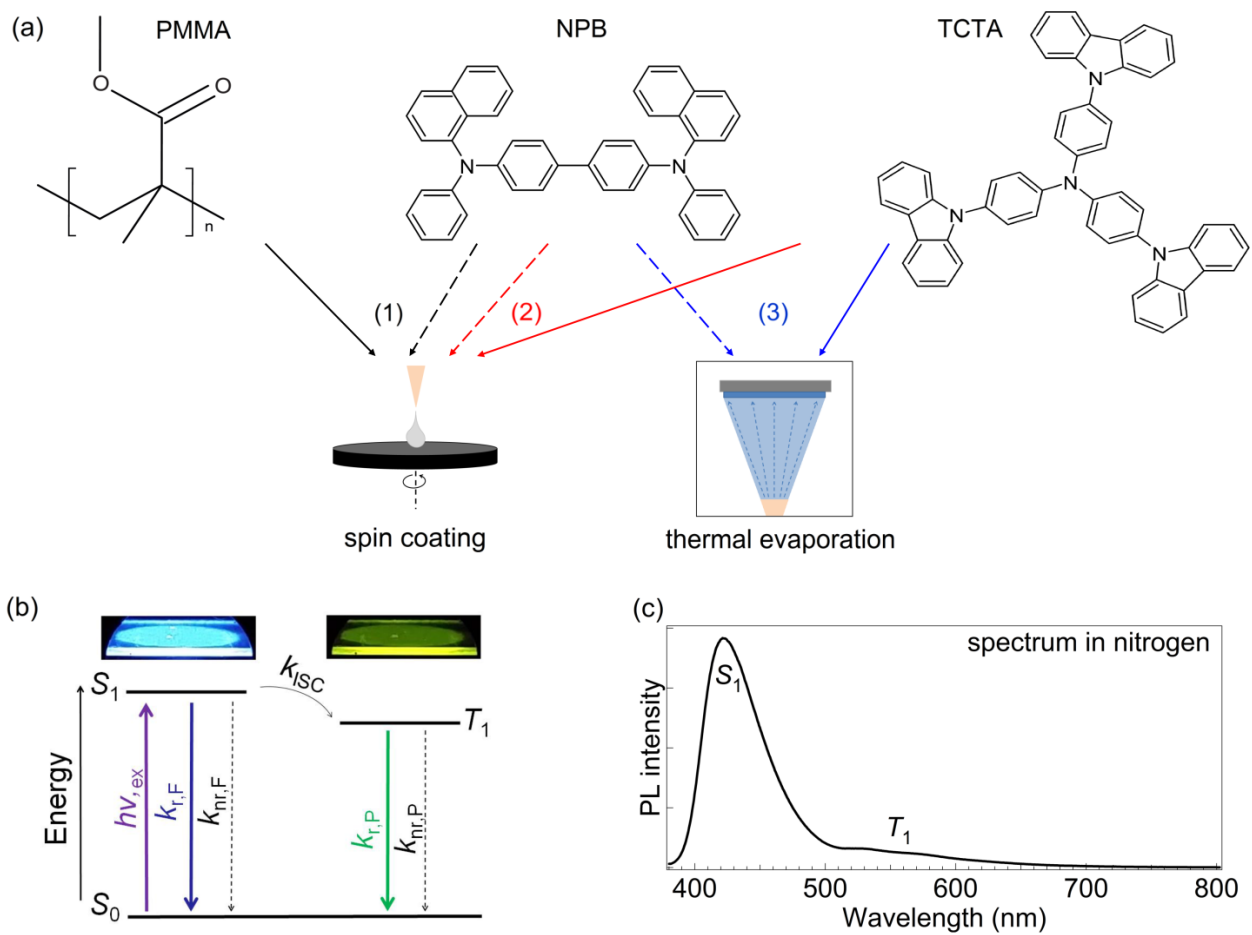

**Figure 1.** (a) From left to right, chemical structures of the polymer matrix PMMA, the biluminophore NPB and the small molecule matrix TCTA. Arrows point out the techniques used to deposit the films as follows: spin coating for PMMA:NPB (black, 1) and TCTA:NPB (red, 2), as well as thermal evaporation for TCTA:NPB (blue, 3). (b) General energy diagram of an organic biluminophore. Above this scheme, a photograph of a spin coated film on a quartz substrate showing the emission from the singlet state ($S_1$) mainly and the afterglow emission from the triplet state ($T_1$). After photon absorption upon excitation ($h\nu_{ex}$) from the ground state ($S_0$), the $S_1$ can be deactivated radiatively as fluorescence ($k_{r,F}$) and non-radiatively via internal conversion ($k_{nr,F}$) or intersystem crossing ($k_{ISC}$) to $T_1$. Following energy transfer from $S_1$, the possible deactivation paths of $T_1$ are radiative as phosphorescence ($k_{r,P}$), and non-radiative ($k_{nr,P}$) via internal conversion and quenching. (c) Photoluminescence (PL) spectra of a typical biluminescent system, under nitrogen atmosphere at room temperature, displaying the contributions from the singlet ($S_1$) and triplet ($T_1$) excited states.

In this study, we present an investigation on the interplay between fluorescence and phosphorescence in such organic biluminophores that puts the consequences of the of the more than six orders of magnitude difference between singlet and triplet states into the spotlight. Here, we vary the host matrices embedding the archetypical biluminescent emitter NPB (*N,N'*-di(1-naphthyl)-*N,N'*-diphenyl-(1,1-biphenyl)-4,4-diamine), which allows for a variation of the triplet population under steady-state conditions as a result of a changed $k_{nr,P}$. Furthermore, we investigate the photoluminescence (PL) under both nitrogen and ambient (air) environments, where in the latter case, the triplet population is effectively quenched by oxygen [19].

For NPB, we have reported that it shows efficient phosphorescence at room temperature (RTP) in addition to conventional fluorescence [9]. Furthermore, NPB is widely known and used in the field of OLEDs as hole transporting material, which is why we use it here as an archetypical biluminescent emitter in this study. It is known that the RTP strongly depends on the host material, as it is the latter that represents the interface to non-radiative deactivation pathways [9-10]. Similar to our earlier reports, we use PMMA with triplet energy of 3.1 eV [20] and glass transition temperature $T_g$ of 108 °C [21] as a reference matrix material based on its ability to form rigid glasses, which is beneficial for observation of phosphorescence at room temperature [9]. Additionally, to vary the environment of the emitter NPB and at the same time resemble the high triplet energy requirement and glassy character of the host in a small molecule, we chose the starburst shaped small molecule TCTA (tris(4-carbazoyl-9-ylphenyl)amine) with triplet energy of 2.79 eV [22] and $T_\text{g}$ of 151 °C [23] as an alternative host material. This material choice is further motivated by the possibility to integrate the effect of biluminescence to electroluminescent devices, which has up to date only been done using specially designed host materials [24]. Figure 1a shows the chemical structures of all three molecules of this study and also illustrates, which processing schemes are used to fabricate the samples. Due to its physical dimensions, the polymer PMMA based system is only fabricated by wet-processing techniques, while for TCTA, the mixed thin films are processed by spin coating (sc) and thermal evaporation. All samples are made with the same NPB concentration of 2 wt% to allow for best comparability. The biluminescence reference system [PMMA:NPB]$_\text{sc}$ 2 wt% shows oxygen dependent phosphorescence, with an excited state lifetime in the range of a few hundred milliseconds (see later discussion for details). The typical fluorescence and phosphorescence emission bands with emission maxima at 425 nm and 540 nm, respectively, are depicted in the biluminescent spectrum plotted in Figure 1c.

Figure 2a shows the time resolved fluorescence and phosphorescence of all samples, which were acquired in the respective time window needed. The corresponding fluorescence and phosphorescence emission bands are plotted for all samples in Figure 2b. The phosphorescence band is detected after the excitation pulse is turned off. All decay curves deviate from monoexponential decay. Therefore, the transient intensities $I$ are fitted with a biexponential decay of the following form (cf. Figure S1):

$$I = A_1 e^{-(t/\tau_1)} + A_2 e^{-(t/\tau_2)} + A_\text{bg}, \qquad (1)$$

Where $A_1$ and $A_2$ are the relative contributions of the individual decays with lifetimes $\tau_1$ and $\tau_2$, respectively. $A_\text{bg}$ is a constant offset that describes the instrument background intensity. It is used in fit calculations, if reached within the time window. Deviation from monoexponential decay, as it is observed here for all systems, has been reported before for similar systems [25]. Average weighted lifetimes for the transients are calculated according to the following expression:

$$\hat{\tau} = \frac{A_1}{A_1+A_2}\tau_1 + \frac{A_2}{A_1+A_2}\tau_2. \qquad (2)$$

These lifetimes are used in the following discussion as representative excited state lifetimes for reasons of simplicity. For all samples, we have made two identical samples and calculated a mean value obtained from the individual samples. The fluorescence lifetimes vary only slightly between the three samples in the range of 2-3 ns (cf. Table 1, values obtained in nitrogen). In stark contrast, the phosphorescence lifetime differs significantly. Here, [PMMA:NPB]$_{sc}$ shows the longest lifetime of 323 ms, followed by the spin-coated [TCTA:NPB]$_{sc}$ system with a lifetime of 96 ms. Clearly the shortest average weighted lifetime is observed for the thermally evaporated (evap) system [TCTA:NPB]$_{evap}$ with 54 ms. Hence, the phosphorescence lifetimes differ by as much a factor of 6 between the [PMMA:NPB]$_{sc}$ reference and the [TCTA:NPB]$_{evap}$ system.

The experimental phosphorescence lifetime can be expressed as:

$$\tau_P = \frac{1}{k_{r,P}+k_{nr,P}} \qquad (3)$$

The change in phosphorescence lifetime cannot be caused by an alteration in the radiative rate $k_{r,P}$, simply because the different chosen environments cannot introduce a coupling strong enough to influence the phosphorescence transition strength. Therefore, the non-radiative rate $k_{nr,P}$ must be responsible for the observed changes. Here, in general, various mechanisms exist that possibly contribute to this rate. While vibrational deactivation $k_{vib}$ is considered to be the primary cause [26], higher order effects like triplet-triplet annihilation ($k_{TTA}$), or even quenching at static impurity sites with a rate $k_{imp}$ can easily add to an effective non-radiative rate. The rate for triplet-triplet annihilation depends on the actual interaction partner concentration:

$$k_{nr,P} = k_{vib} + k_{TTA}(n_T) + k_{imp}. \qquad (4)$$

While the intermolecular forces for small molecules are the same as for polymers (i.e. van der Waals forces), polymer molecules are large systems, where the magnitude of their intermolecular forces can exceed those between small molecules [27]. Therefore, a polymer matrix like PMMA brings higher rigidity to embedded NPB molecules, which in turn suppresses non-radiative deactivation [9,10,28]. For the difference between the two TCTA based systems (sc and evap), we can only speculate at this stage. It is possible that the formation by spin coating, being a highly non-equilibrium process, induces mechanical constrains to the system that similarly introduces rigidity to the film that suppresses non-radiative channels. In contrast, the fabrication via thermal evaporation will allow the molecules to find their energetically most favorable conformation within the mixed film, which leads to a softer packing allowing for more non-radiative modes. We have investigated these systems systematically with X-ray diffraction and Fourier-transform infrared spectroscopy (cf. Figure S2) to possibly identify differences in the constitution of the different samples that may hint to the arguments from above. However, we did not observe significant differences. All samples are amorphous, allowing us to exclude effects of crystal formation to affect our observations, which is an alternative route to enhance RTP [29-31]. We have observed delayed fluorescence as a result of TTA. Here, the large singlet-triplet energy splitting of NPB (~ 0.6 eV [32]) rules out up-conversion via thermally activation, which can only overcome barriers of less than 0.37 eV [33]. TTA-derived delayed fluorescence has been previously reported only for organic crystals exhibiting room temperature phosphorescence [34] as a side effect of the proximity between molecules and formation of agglomerates. In the three biluminescent systems, TTA is strongest for the [TCTA:NPB]$_{evap}$ samples (cf. Figure S3). For the remaining discussion, we consider non-radiative losses as one effective rate $k_{nr,P}$ as given in Equation (4) influencing the phosphorescent lifetime $\tau_p$, knowing that it, strictly seen, is a quantity that decreases during the excited state decay, if TTA is significant.

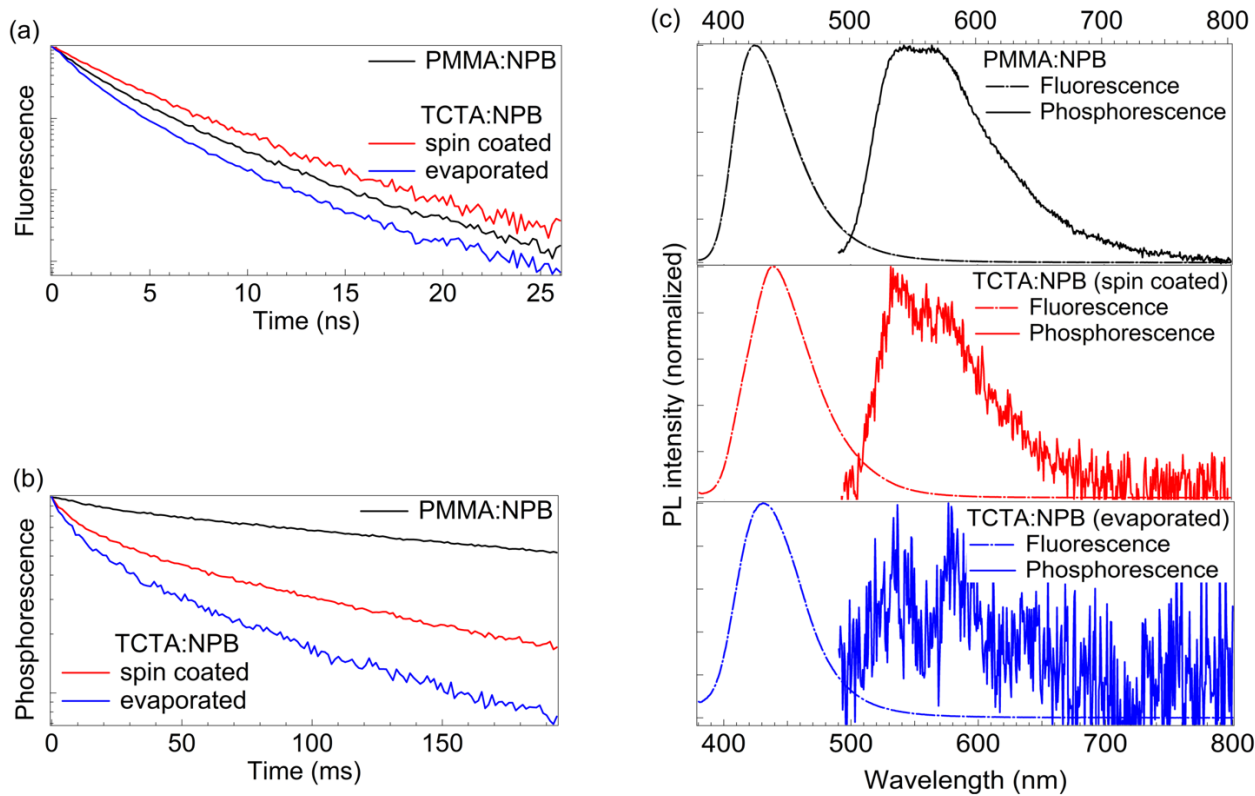

**Figure 2.** Characteristics of biluminescent systems when the emitting molecule (NPB) is diluted at 2 wt% in the host matrix. (a) Normalized dynamic response under nitrogen conditions in the nanosecond range, corresponding to the prompt fluorescence radiative decay rate. (b) Normalized dynamic response under nitrogen conditions in the millisecond range, corresponding to the delayed (phosphorescence) radiative decay rate. (c) Normalized fluorescence spectra (dashed line) of biluminescent films excited at $\lambda_{ex}$ = 365 nm and continuous wave source, and phosphorescence spectra (solid line) of biluminescent films excited at $\lambda_{ex}$ = 365 nm with a pulsed source, under nitrogen atmosphere at room temperature. Residual contributions from delayed fluorescence are cut in the spectra of (c). For the complete delayed emission spectra, refer to Figure S3.

Possible applications making use of biluminescence such as optical sensing, do require operation in quasi-continuous wave (cw) regime to allow the phosphorescence intensity to build up properly. Here, quasi-cw is understood as a scenario where the pump duration is in the same order of magnitude as the triplet state lifetime $\tau_p$. For a molecule similar to NPB, namely (BzP)PB [*N,N'*-bis(4-benzoyl-phenyl)-*N,N'*-diphenyl-benzidine], we have already observed a decrease of the integrated luminescence, i.e. fluorescence and phosphorescence, during a quasi-cw excitation pulse [25]. While appointing this effect to singlet-triplet annihilation (STA) with a rate $k_{STA}$ [35], we did not investigate this effect further [25].

This important interplay between the two luminescent states will be discussed in the following. The strength of STA is proportional to both singlet $n_S$ and triplet exciton densities $n_T$ [35]:

$$\left.\frac{dS}{dt}\right|_{STA} = k_{STA} n_S n_T \qquad (5)$$

With the three systems discussed above having distinctively different triplet lifetimes $\tau_p$ (cf. Table 1), the magnitude of STA should also vary between the samples, as the triplet exciton density $n_T$ scales with the lifetime $\tau_p$. To reduce STA to an absolute minimum, we performed additional measurements (transient and quasi-cw) in ambient environment (air), making use of the fact that oxygen very effectively quenches the triplet states with a rate $k_{q,O_2}$ [19, 26]. Hence, under the presence of oxygen, the Equation (3) becomes:

$$\tau_P = \frac{1}{k_{r,P}+k_{nr,P}+k_{q,O_2}}. \qquad (6)$$

Figure 3 shows the fluorescence (a) and phosphorescence (b) decays following a pulsed excitation for all three systems. The fluorescence shows no differences between nitrogen and air measurements. This result with the additional information that all three systems show slightly different fluorescence decays (cf. Figure 2a and Table 1) lead to the conclusion that none of the three involved rates, i.e. $k_{r,F}$, $k_{nr,F}$, and $k_{ISC}$, are altered by the presence of oxygen. In contrast, the phosphorescence decays of all samples are much faster in air compared to the measurement in nitrogen atmosphere, as a result of an effective $k_q$. Interestingly, all three transients in air show very similar time constants showing some residual radiative contribution giving rise to a $\tau_p \sim 10$ ms (cf. Table 1). This suggests that the oxygen quenching is strong but not completely outcompeting $k_{r,P}$. For all samples, the oxygen quenching effectively reduces the triplet density under steady-state conditions by at least one order of magnitude (cf. Table 1).

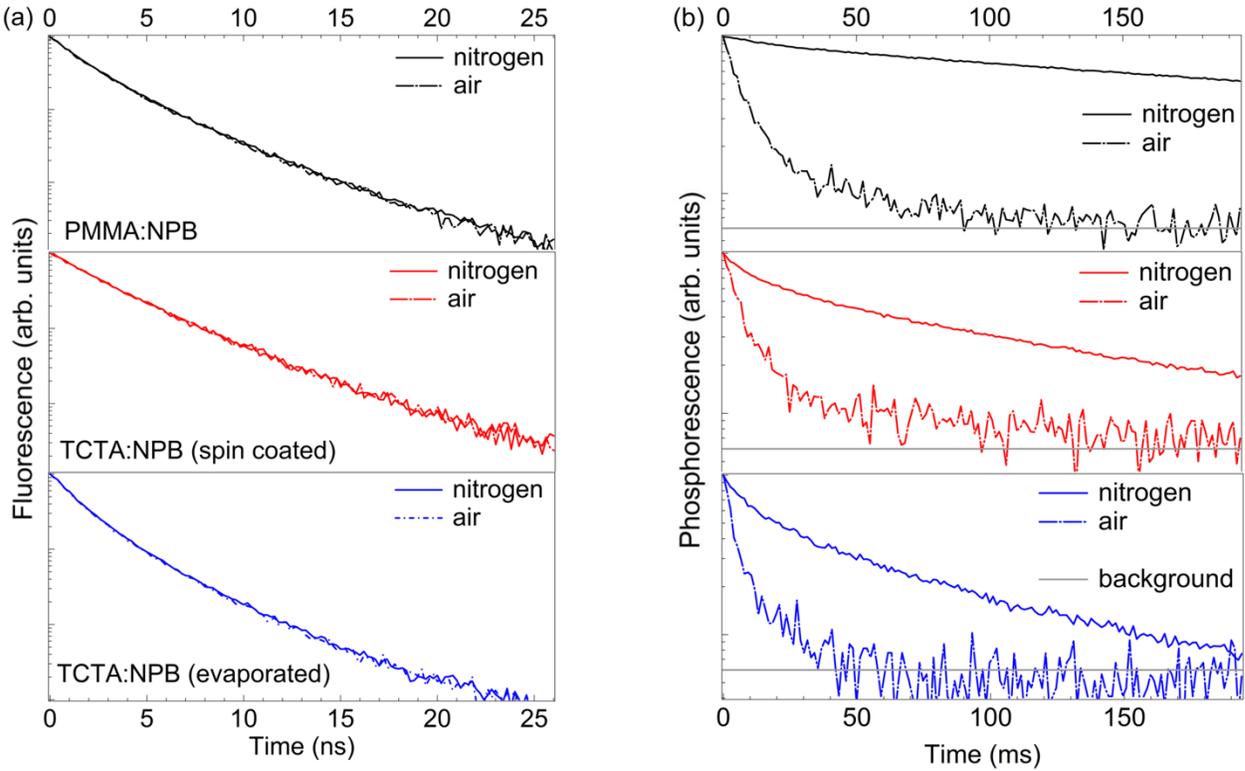

**Figure 3.** Time resolved luminescence of the different biluminescent systems [PMMA:NPB]$_{sc}$ (black), [TCTA:NPB]$_{sc}$ (red), and [TCTA:NPB]$_{evap}$ (blue) in nitrogen (solid line) and air (dashed line). (a) Fluorescence decays following a single, sub-ns pulse with $\lambda_{ex} = 374$ nm. (b) Phosphorescence decays of the samples following a high repetition (80 MHz) burst excitation composed of many sub-ns pulses ($\lambda_{ex} = 374$ nm). This is to assure a sufficiently high triplet

population to detect the long-lived phosphorescence. The horizontal grey lines indicate the instrument background.

**Table 1.** Photophysical properties of biluminescent systems at room temperature.

| System [a] | $F_{max}$ [nm][b] | $P_{max}$ [nm][c] | $\tau_F$[ns][d] $N_2$ | $\tau_F$[ns][d] air | $\tau_P$[ms][e] $N_2$ | $\tau_P$[ms][e] air | PLQY [%][f] | PLQY$_F$ [%][g] | PLQY$_P$ [%][h] | $I_{F,N_2}/I_{F,O_2}$ [%][i] |
|---|---|---|---|---|---|---|---|---|---|---|
| [PMMA:NPB]$_{sc}$ | 425 | 540 | 2.6 | 2.6 | 322.6 | 10.4 | 26±3 | 23.1 | 2.6 | 77 |
| [TCTA:NPB]$_{sc}$ | 440 | 540 | 3.4 | 3.4 | 95.6 | 10.1 | 31±8 | 30.4 | 0.4 | 96 |
| [TCTA:NPB]$_{evap}$ | 430 | 540 | 2.0 | 2.3 | 53.8 | 7.1 | 32±3 | 31.6 | 0.2 | 98 |

[a] Data for systems in which NPB is diluted at 2 wt% in the host matrix; measured fluorescence [b] and phosphorescence [c] emission maxima; fitted (biexponential, average weighted) fluorescence [d] and phosphorescence [e] lifetimes; photoluminescence quantum yield [f]; fluorescence quantum yield [g]; phosphorescence quantum yield [h]; nitrogen-to-air fluorescence intensity maxima ratio [i].

Figure 4a-c show the cw-PL of all samples obtained under exactly the same measurement configuration for both nitrogen and ambient environments. In air, the remaining phosphorescence contribution (cf. Figure 3b) is so weak that it is masked by the low energy tail of the fluorescence spectrum. The PL spectra in nitrogen atmosphere show two important differences: (i) the relative phosphorescence intensity scales with the phosphorescence lifetimes, where the reference system [PMMA:NPB]$_{sc}$ shows the highest phosphorescence intensity. (ii) The fluorescence intensity of all samples decreases with respect to the measurement in air. As mentioned above at the discussion of the time resolved data, none of the outgoing rates from the $S_1$ level can be the reason for this observation. Figure 4d shows the ratio between peak fluorescence under nitrogen and air (also cf. Table 1), showing the strongest reduction for the [PMMA:NPB]$_{sc}$ system (77%) and the weakest

one for the [TCTA:NPB]$_{evap}$ (98%) – again in agreement with STA scaling with the triplet exciton density $n_T$.

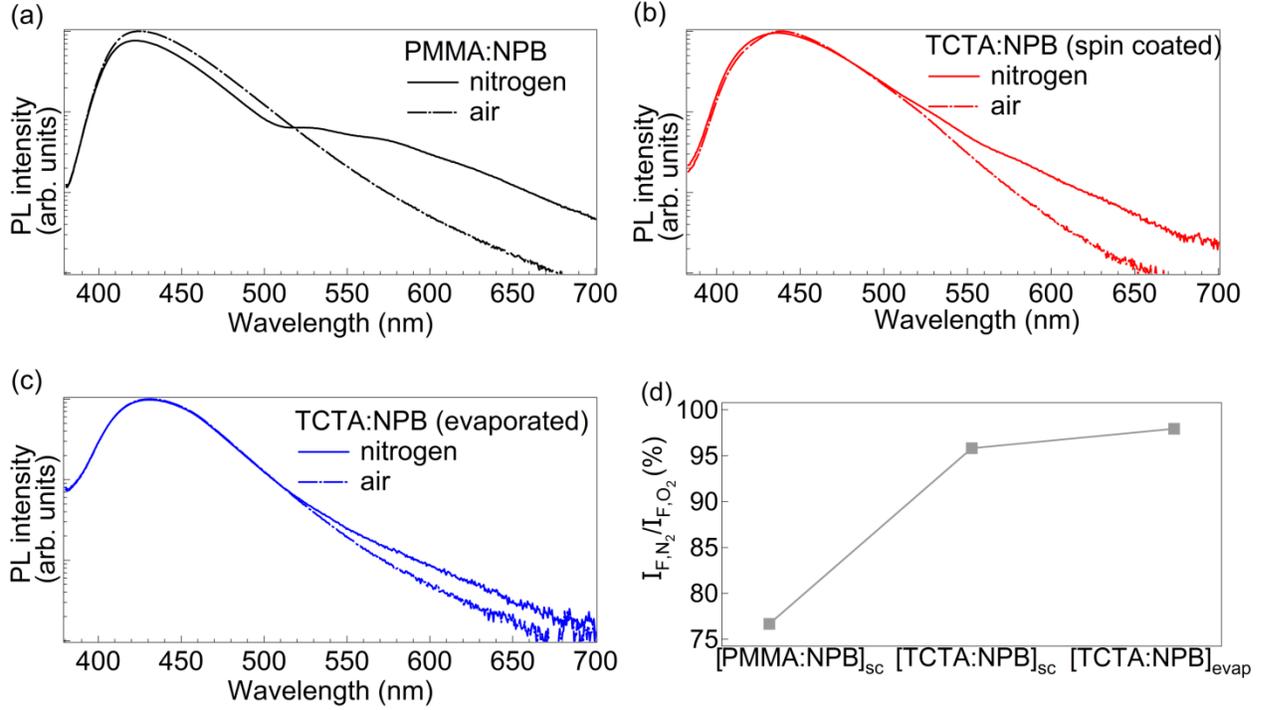

**Figure 4.** Comparison of the spectral characteristics of the different samples in nitrogen and air. Photoluminescence spectra are taken under continuous wave excitation, in both air (dashed line) and nitrogen (solid line) environments without changing the geometry of the setup, allowing for comparability of the respective two measurements. Spectra are normalized with respect to the fluorescence maximum to the luminescence in air. (a) [PMMA:NPB]$_{sc}$ (black), (b) [TCTA:NPB]$_{sc}$ (red), (c) [TCTA:NPB]$_{evap}$ (blue), (d) ratio of the absolute fluorescence peaks between nitrogen and air environment for the three different systems [(a)-(c)].

Figure 5a shows the PL intensity as a function of time for the [PMMA:NPB]$_{sc}$ reference system during 500 ms on, 500 ms off duty cycle quasi-cw pulse. While the air measurement shows constant intensity during the on-pulse, the corresponding intensity obtained under nitrogen atmosphere gradually decreases reaching a steady state close to the off-time. Figure 5b shows the intensity of the fluorescence and phosphorescence as a function of time during the on-pulse as obtained from spectrally resolved time-gated measurements. Here, the fluorescence decreases while at the same time, the phosphorescence increases. For the phosphorescence data points, it is important to note that the spectral integration window (500 – 700 nm) accounts not only for the phosphorescence, but also for the low energy tail of the fluorescence band (cf. Figure 4a). Therefore, the increase of the phosphorescence is counteracted by a fraction of the fluorescence, which decreases with time. A detailed analysis of the phosphorescence during the on-pulse is given in the supporting information (cf. Figure S5). The increase in phosphorescence, and by that the triplet exciton density $n_T$, is due to the long phosphorescent lifetime $\tau_p \sim 330$ ms of the [PMMA:NPB]$_{sc}$ system. In accordance with Equation (4), the singlet exciton density $n_S$, and consequently the fluorescence intensity, will decrease with increasing $n_T$, as excited singlet states are absorbed by excited triplet states with a rate $k_{STA}$. This process is an inherent loss channel for the biluminescent system, as singlet exciton energy is dissipated in a non-radiative $T_1 \xrightarrow{STA} T_n \xrightarrow{relaxation} T_1$ cycle. This is also reflected in the PLQY data of the three systems. While the two systems comprising TCTA as host, which show only minor reductions in fluorescence between air and nitrogen measurement (cf. Figure 4d), show 31% and 32% PLQY for [TCTA:NPB]$_{sc}$ and [TCTA:NPB]$_{evap}$, respectively, the corresponding value for the [PMMA:NPB]$_{sc}$ reference systems is lower (26%). This difference in PLQY fits well to the reduction of 77% observed in Figure 4d (cf. Table 1).

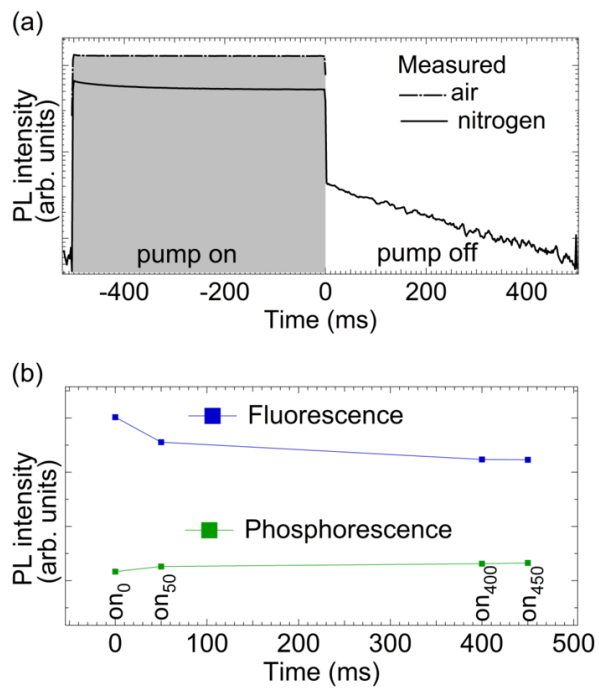

**Figure 5.** Development of singlet and triplet populations during quasi-cw excitation pulses. (a) PL intensity of the [PMMA:NPB]$_{sc}$ reference system during and after a 500 ms excitation pulse under air (dashed line) and nitrogen (solid line) environments. (b) Intensities of fluorescence (380 – 500 nm) and phosphorescence (500 – 700 nm) obtained in 20 ms integration windows with different delays to the pump pulse start. Here, on$_0$ = no delay, on$_{50}$ = 50 ms delay, on$_{400}$ = 400 ms delay, and on$_{450}$ = 450 ms delay.

In conclusion, this study discusses the exciton dynamics of the efficient, room temperature biluminescent emitter NPB. The radiative lifetimes of fluorescence and phosphorescence are in the range of 3 ns and 300 ms, respectively, spanning eight orders of magnitude. This difference in lifetimes puts the biluminescent system in strong imbalance between singlet and triplet exciton densities under continuous wave operation, which can lead to a limitation of the biluminescence efficiency as a result of non-linear processes. Here, singlet-triplet annihilation (STA) reduces the fluorescence intensity significantly. By sample-engineering, the triplet exciton lifetime can be reduced through an increase of the non-radiative rate $k_{nr,P}$, which leads to an effectively suppressed STA. Hence, depending on the actual pump intensities that will dictate the exciton densities in both spin manifolds, an upper limit for the phosphorescence lifetime can be deduced until which STA is only of minor effect. Moreover, a convenient and simple method to observe biluminescence from purely organic molecules is demonstrated, in which commercially available small molecules and polymer are mixed to form a biluminescent host-guest system, with no need of complicated material synthesis or functionalization. Our report shows in addition that small molecule matrix materials, which normally diverge significantly in their nanostructure from polymers, are also capable of promoting biluminescence of purely organic molecules at room temperature.

While for some applications, the persistent nature of the room temperature phosphorescence is a key feature, other areas of use may exclusively call for high efficiency and broadband, dual state emission. Future material design for biluminescent or RTP emitters will have a careful optimization of the triplet state's dynamics to be of central importance.

EXPERIMENTAL SECTION

The studies were carried out on thin films deposited either by spin coating or thermal evaporation in ultrahigh vacuum, respectively. The biluminescent emitter is NPB (*N,N'*-di(1-naphthyl)-*N,N'*-diphenyl-(1,1-biphenyl)-4,4-diamine) (Lumtec), which is dispersed into either the polymer host PMMA (Alfa Aesar) or the small molecule host TCTA (Lumtec). PMMA was used as received, TCTA and NPB were purified by vacuum train sublimation twice before use. Solvent polarity and dilution level were optimized for the wet processing of the small molecule host system (TCTA), in order to obtain homogeneous solutions and films [36]. Here, all materials were dissolved in methoxybenzene (Anisol, purity 99.7 %) at concentrations of 80, 40 and 10 mg/mL for PMMA, TCTA and NPB, respectively. Quartz substrates were cleaned based on the following sequence: 15 minutes treatment in ultrasonic bath ($UB_{15}$) in DIW and soap; rinsing with DIW; $UB_{15}$ in ethanol, $UB_{15}$ in IPA and finally 10 min of $O_2$ plasma after drying with nitrogen. For the spin coated films, NPB is embedded at a concentration of 2 wt % onto the host (PMMA or TCTA). Such dilutions are spin coated from a methoxybenzene solution onto glass substrates (1" by 1" by 1 mm) at a spin speed of 2000 rpm (1000 rpm/s ramp) for 60 s. Thermal evaporated films were formed by co-evaporation of TCTA and NPB on an unheated quartz glass substrate, using an ultra-high vacuum deposition system (Kurt J. Lesker Company) at a deposition rate of 0.8 A/s under a pressure below $3 \cdot 10^{-7}$ mbar. The target concentration was 2 wt% as well to match the solution processed samples.

PL measurements were performed under nitrogen atmosphere ($N_2$ 6.0, purity 99.9999 %) and ambient conditions (air) at room temperature. For the acquisition of time resolved phosphorescence and emission spectra data, a customized set up was used, where nitrogen is flowing continuously in a sample box to guarantee an inert atmosphere during measurements. PL spectra were recorded with a spectrometer (CAS 140CTS, Instrument Systems) and the LED (M365L2, $\lambda_{max}$ =365 nm, Thorlabs) operating in continuous wave mode to obtain the integrated spectrum. The LED was driven by a T-Cube LED driver (LEDD1B, Thorlabs), a band-pass filter (FB370-10, Thorlabs) was placed in front of the LED to avoid an overlap between excitation and emission spectra. Phosphorescence spectra were collected with the same instruments with the addition of the pulse generator (Keysight (Agilent) Technologies 8114A) to separate phosphorescence from fluorescence in a time-gated scheme. Therefore, a pulse train of 500 ms

LED-on and 500 ms LED-off was generated by the pulse generator. The spectrometer was triggered with a delay of 20 ms to the falling edge of the LED-on pulse and integrated for 450 ms, while the excitation source was off, assuring no detection of the next pulse. The spectrometer CAS 140CTS was gated to 20 ms integration windows, to resolved the PL during a 500 ms long pump pulse of the UV LED. These windows were started with different delays (0, 50… 450 ms) to the beginning of the pump pulse. The time resolved PL intensity of a complete duty cycle was recorded with a switchable gain silicon photodetector (PDA100A, Thorlabs).

The time resolved PL setup to measure the fluorescence and phosphorescence lifetimes, consisted of a time correlated single photon counting (TCSPC) system (TimeHarp 260 PICO, Picoquant). The excitation source was a pulsed diode laser PDL 820 with a 374 nm laser head running at 40 MHz repetition rate. A single photon sensitive PMA Hybrid was used for detection. The count rate was adjusted to < 1 % of the laser rate to prevent pile-up. Data acquisition was set to a resolution of 0.2 ns to record the prompt fluorescence lifetime and 1.3 ms for the phosphorescence lifetime. Phosphorescence decays were recorded following a high repetition (80 MHz) burst excitation composed of many sub-ns pulses ($\lambda_{ex}$ =374 nm), which, for the long phosphorescence lifetime, functions as a high intensity excitation pulse.

Photoluminescence quantum yield (PLQY) measurements were performed in a customized setup for measurements under nitrogen atmosphere at oxygen levels below 0.1%. This setup consists of an antistatic glove box with transfer chamber (SICCO), 6" integrating sphere (Labsphere) using a LED (M340L4, $\lambda_{max}$ =340 nm, Thorlabs) as excitation source, driven by the LEDD1B driver (Thorlabs) and a spectrometer (CAS 140CTS, Instrument Systems), following de Mello method [37] for the estimation of the absolute PLQY, and the spectra were corrected for wavelength-dependent instrument sensitivity.

The thin films' morphology was analyzed by X-ray reflectivity (XRR). The mean film density was determined from a 13 point X-ray reflectivity (XRR) measurement, using a Jordan Valley JVX 5200 XRR thin film measurement system. The characterization of the vibrational modes was performed using a Nicolet™ iS™ 10 FT-IR Spectrometer (Thermo Fisher Scientific) in attenuated total reflection (ATR) mode, with a diamond crystal was used as window and a 20nm gold layer underneath the film of interest.

ASSOCIATED CONTENT

**Supporting Information**

Additional data of the fluorescence and phosphorescence transients and curve fittings; Fourier transform infrared spectroscopy (FTIR), X-ray reflectivity (XRR) and grazing incidence X-ray diffraction (GIXRD) spectra of the TCTA:NPB system; additional data of the photoluminescence spectra showing TTA-based delayed fluorescence in PMMA:NPB and TCTA:NPB systems; calculation of the fluorescence and phosphorescence quantum yield; experimental determination of the singlet and triplet populations of the PMMA:NPB system. This material is available free of charge via the Internet at http://pubs.acs.org

AUTHOR INFORMATION

**Notes**

The authors declare no competing financial interests.

ACKNOWLEDGEMENTS

The authors are grateful for the support of the German Excellence Initiative via the Cluster of Excellence EXC 1056 Center for Advancing Electronics Dresden (cfaed). This work has received funding from the European Research Council (ERC) under the European Union's Horizon 2020 research and innovation programme (grant agreement No 679213). The authors thank Caroline Walde, Tobias Günther, and Andreas Wendel for film evaporation, Dr. Lutz Wilde from Fraunhofer IPMS for XRR measurements, Dr. Norwid-Rasmus Behrnd and Dr. Ramunas Lygaitis for discussions concerning the morphology of the films, and finally, Dr. Axel Fischer and Dr. Reinhard Schloss for further discussions on the topic of this manuscript.

[14] Gong, Y., et al. Achieving persistent room temperature phosphorescence and remarkable mechanochromism from pure organic luminogens. Adv. Mater. 27, 6195–6201 (2015).

[15] Xu, J., Takai, A., Kobayashia, Y., Takeuchi, M. Phosphorescence from a pure organic fluorene derivative in solution at room temperature. Chem. Commun. 49, 8447–8449 (2013).

[16] Li, C., et al. Reversible luminescence switching of an organic solid: controllable on–off persistent room temperature phosphorescence and stimulated multiple fluorescence conversion. Adv. Opt. Mater. 3, 1184–1190 (2015).

[17] Hirata, S. et al. Large reverse saturable absorption under weak continuous incoherent light. Nat. Mater. 13, 938–946 (2014)

[18] Lehner, P., Staudinger, C., Borisov, S. M., Klimant, I. Ultra-sensitive optical oxygen sensors for characterization of nearly anoxic systems. Nat. Comm. 5, 4460 (2014)

[19] Garner, A., Wilkinson, F. Quenching of triplet states by molecular oxygen and the role of charge-transfer interactions. Chem. Phys. Lett. 45, 432–435 (1977).

[20] Avdeenko, A. A., Dobrovolskaya, T. L., Kultchitsky, V. A., Nabiokin, Y. V. Pakulov, S. N. Temperature-dependence of luminescence decay time of benzyl. J. Lumin.11, 331–337 (1976).

[21] Porter, C. E. & Blum, F. D. Thermal characterization of PMMA thin films using modulated differential scanning calorimetry. Macromolecules. 33, 7016-7020 (2000).

[22] Reineke, S., Schwartz, G., Walzer, K., Leo, K. Direct observation of host-guest triplet-triplet annihilation in phosphorescent solid mixed films. Phys. Stat. Sol. RRL3, 67–69 (2009).

# Supporting Information

# Interplay of Fluorescence and Phosphorescence in Organic Biluminescent Emitters

*Caterin Salas Redondo,[1,2] Paul Kleine,[1] Karla Roszeitis,[1] Tim Achenbach,[1] Martin Kroll,[1] Michael Thomschke,[1] and Sebastian Reineke*[1,2]*

[1] Dresden Integrated Center for Applied Physics and Photonic Materials (IAPP) and Institute for Applied Physics, Technische Universität Dresden, Nöthnitzer Straße 61, D-01187, Germany

[2] Center for Advancing Electronics Dresden (cfaed), Technische Universität Dresden, Würzburger Straße 46, D-01187 Dresden, Germany

AUTHOR INFORMATION

**Corresponding Author**

*Sebastian Reineke. E-mail: reineke@iapp.de

CONTENT

1. Detailed calculation of the photoluminescence transients

2. Morphology analysis of the biluminescent system TCTA:NPB

3. TTA-based delayed fluorescence

4. Calculation of the fluorescence and phosphorescence quantum yield

5. Experimental determination of the singlet and triplet population of the biluminescent system PMMA:NPB

# 1. Detailed calculation of the photoluminescence transients

For each system composition (i.e. [PMMA:NPB]sc, [TCTA:NPB]sc and [TCTA:NPB]evap), two identical samples are made, which are fitted independently using:

$$I = A_1 e^{-(t/\tau_1)} + A_2 e^{-(t/\tau_2)} + A_{bg} \qquad (1)$$

Where $A_1$ and $A_2$ are the relative contributions of the individual decays with lifetimes $\tau_1$ and $\tau_2$, respectively. $A_{bg}$ is a constant offset that describes the instrument background intensity.

Afterwards, the average weighted lifetimes $\hat{\tau}$ are calculated according to:

$$\hat{\tau} = \frac{A_1}{A_1+A_2}\tau_1 + \frac{A_2}{A_1+A_2}\tau_2. \qquad (2)$$

Finally, the mean value of the lifetimes $\hat{\tau}_{av}$ are calculated out of the results of the two samples per system. The final results are given here as well as in the main text (cf. Table 1). Moreover, we present the values of all the parameters obtained after the fitting process to find both, the fluorescence and phosphorescence lifetimes.

Data corresponding to the fitting of fluorescence lifetime:

[PMMA:NPB]sc

| Sample | Atmosphere | $A_{bg}$ | $A_1$ | $\tau_1$ (ns) | $A_2$ | $\tau_2$ (ns) | $\hat{\tau}$ (ns) | $\hat{\tau}_{av}$ (ns) |
|---|---|---|---|---|---|---|---|---|
| A | Nitrogen | 0 | 0.67 | 1.8 | 0.34 | 4.3 | 2.6 | 2.6 |
| B |  | 0 | 0.67 | 1.8 | 0.34 | 4.3 | 2.6 |  |
| A | Air | 0 | 0.70 | 1.6 | 0.31 | 4.1 | 2.4 | 2.6 |

| Sample | Atmosphere | $A_{bg}$ | $A_1$ | $\tau_1$ (ns) | $A_2$ | $\tau_2$ (ns) | $\hat{\tau}$ (ns) | $\hat{\tau}_{av}$ (ns) |
|---|---|---|---|---|---|---|---|---|
| B | | 0 | 0.59 | 1.7 | 0.43 | 4.2 | 2.7 | |

[TCTA:NPB]$_{sc}$

| Sample | Atmosphere | $A_{bg}$ | $A_1$ | $\tau_1$ (ns) | $A_2$ | $\tau_2$ (ns) | $\hat{\tau}$ (ns) | $\hat{\tau}_{av}$ (ns) |
|---|---|---|---|---|---|---|---|---|
| C | Nitrogen | 0 | 0.62 | 4.0 | 0.39 | 1.7 | 3.1 | 3.4 |
| D | | 0 | 0.92 | 3.1 | 0.12 | 7.4 | 3.6 | |
| C | Air | 0 | 0.72 | 2.5 | 0.28 | 4.9 | 3.2 | 3.4 |
| D | | 0 | 0.76 | 2.8 | 0.24 | 5.9 | 3.5 | |

[TCTA:NPB]$_{evap}$

| Sample | Atmosphere | $A_{bg}$ | $A_1$ | $\tau_1$ (ns) | $A_2$ | $\tau_2$ (ns) | $\hat{\tau}$ (ns) | $\hat{\tau}_{av}$ (ns) |
|---|---|---|---|---|---|---|---|---|
| E | Nitrogen | 0 | 0.25 | 3.8 | 0.76 | 1.4 | 2.0 | 2.0 |
| F | | 0 | 0.78 | 1.5 | 0.23 | 3.6 | 2.0 | |
| E | Air | 0 | 0.27 | 3.7 | 0.74 | 1.4 | 2.0 | 2.3 |
| F | | 0 | 0.24 | 4.2 | 0.77 | 2.0 | 2.5 | |

On the same regard, the complete data obtained from the fitting of the phosphorescence is then:

[PMMA:NPB]$_{sc}$

| Sample | Atmosphere | $A_{bg}$ | $A_1$ | $\tau_1$ (ms) | $A_2$ | $\tau_2$ (ms) | $\hat{\tau}$ (ms) | $\hat{\tau}_{av}$ (ms) |
|---|---|---|---|---|---|---|---|---|
| A | Nitrogen | 0 | 0.10 | 17.5 | 0.9 | 382.2 | 346.3 | 322.6 |
| B | | 0 | 0.12 | 22.0 | 0.88 | 337.2 | 298.8 | |
| A | Air | 0.06 | 0.83 | 7.4 | 0.11 | 34.6 | 10.7 | 10.4 |
| B | | 0.06 | 0.75 | 5.3 | 0.19 | 28.9 | 10.0 | |

[TCTA:NPB]$_{sc}$

| Sample | Atmosphere | $A_{bg}$ | $A_1$ | $\tau_1$ (ms) | $A_2$ | $\tau_2$ (ms) | $\hat{\tau}$ (ms) | $\hat{\tau}_{av}$ (ms) |
|---|---|---|---|---|---|---|---|---|
| C | Nitrogen | 0 | 0.36 | 12.5 | 0.64 | 162.4 | 108.6 | 95.6 |
| D |  | 0 | 0.38 | 11.3 | 0.62 | 125.9 | 82.5 |  |
| C | Air | 0.06 | 0.84 | 6.0 | 0.10 | 46.1 | 10.2 | 10.1 |
| D |  | 0.06 | 0.85 | 6.3 | 0.09 | 45.6 | 10.0 |  |

[TCTA:NPB]$_{evap}$

| Sample | Atmosphere | $A_{bg}$ | $A_1$ | $\tau_1$ (ms) | $A_2$ | $\tau_2$ (ms) | $\hat{\tau}$ (ms) | $\hat{\tau}_{av}$ (ms) |
|---|---|---|---|---|---|---|---|---|
| E | Nitrogen | 0 | 0.44 | 11.0 | 0.56 | 102.8 | 62.0 | 53.8 |
| F |  | 0 | 0.53 | 9.9 | 0.47 | 86.0 | 45.5 |  |
| E | Air | 0.2 | 0.47 | 13.3 | 0.53 | 2.2 | 7.5 | 7.1 |
| F |  | 0.05 | 0.13 | 25.1 | 0.87 | 4.0 | 6.7 |  |

Furthermore, a set of decay curves that represent the fluorescence and phosphorescence transients under air and nitrogen atmospheres, with their equivalent curve fittings to align theoretical curves with normalized experimental transients are given below:

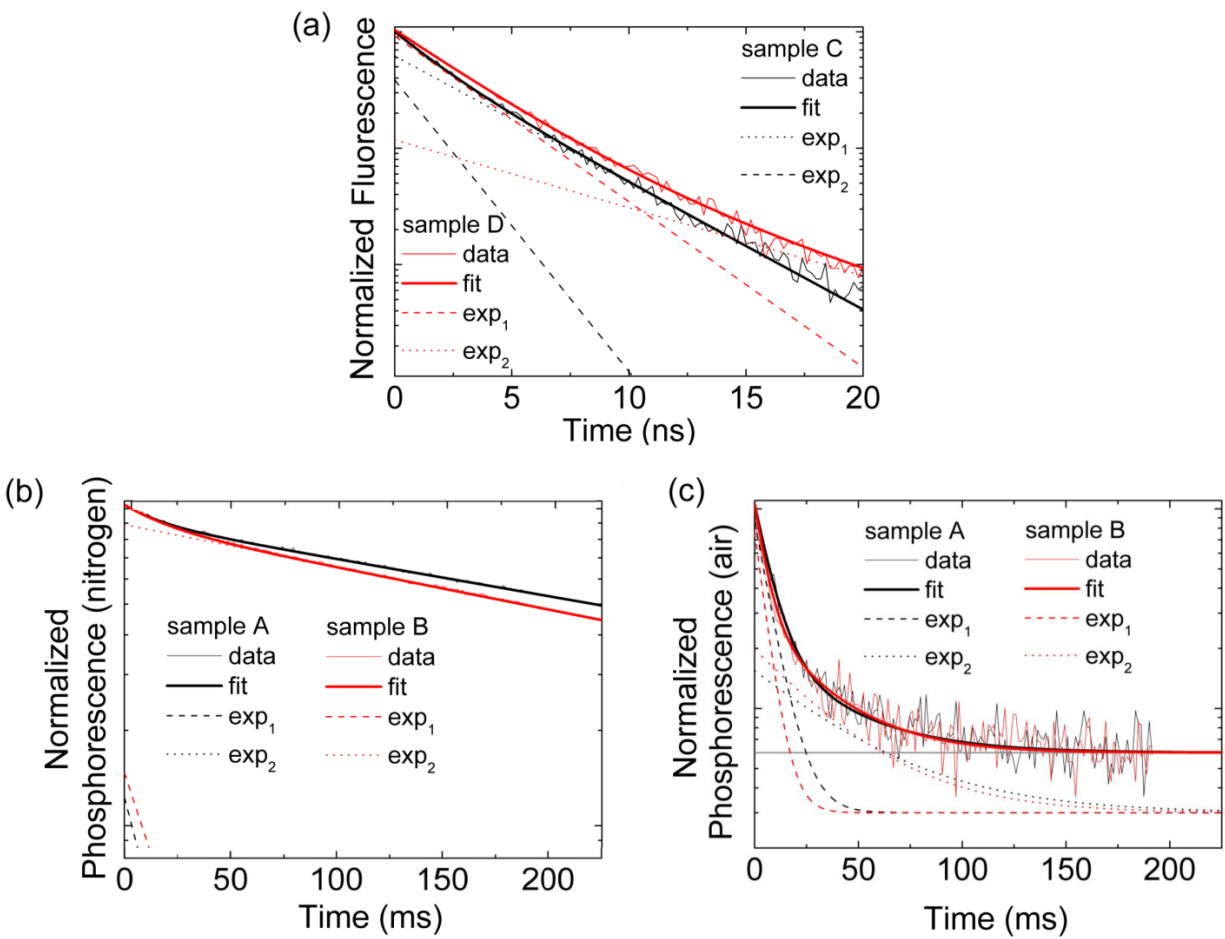

**Figure S1.** Typical double exponential fitting curves of the transient response. Solid bold line represents the double exponential fit, dashed and dotted lines correspond to the two monoexponential components $exp_1$ and $exp_2$ respectively. (a) Fit of the fluorescence transient of [TCTA:NPB]$_{sc}$ samples C (black) and D (red). (b) Fit of the phosphorescence transient in nitrogen of [PMMA:NPB]$_{sc}$ samples A (black) and B(red). (c) Fit of the phosphorescence transient in air of [PMMA:NPB]$_{sc}$ samples A (black) and B (red), the grey line represents the background level of the measurement system.

## 2. Morphology analysis of the biluminescent system TCTA:NPB

We have investigated the morphology of the biluminescence systems [TCTA:NPB]$_{sc}$ and [TCTA:NPB]$_{evap}$ with X-ray diffraction and Fourier transform infrared spectroscopy, in order to determine whether there are significant differences affecting the exciton dynamics of the biluminescent emitter NPB, depending on the technique we used to deposit the films.

Figures S2a and S2b are dedicated to the X-ray-reflectometry (XRR) and Grazing incidence X-ray diffraction (GIXRD) of the films, respectively. The density of thin films can be estimated from the critical angle $\theta_c$ for total reflection of the signal in the XRR method [1] (cf. Figure S2a). Here, a shift in the total reflection edge is not apparent between both XRR signals, which implies that if at all, there is no significant change in density. Although the film density yielded to similar estimation, this does not suggest a similar local rigidity around the NPB molecules.

The GIXRD spectrum was obtained with incidence angles of 0.2095° and 0.184° for the TCTA:NPB spin coated and evaporated samples respectively. The diffraction peaks are not sharp for both vacuum-deposited and spin-coated films (cf. Figure S2b), indicating that no crystallization is present in the films [2] but rather they describe the profile of a disorder lattice of amorphous materials. This leads us to the conclusion that no aggregation or crystallization was induced in none of the samples.

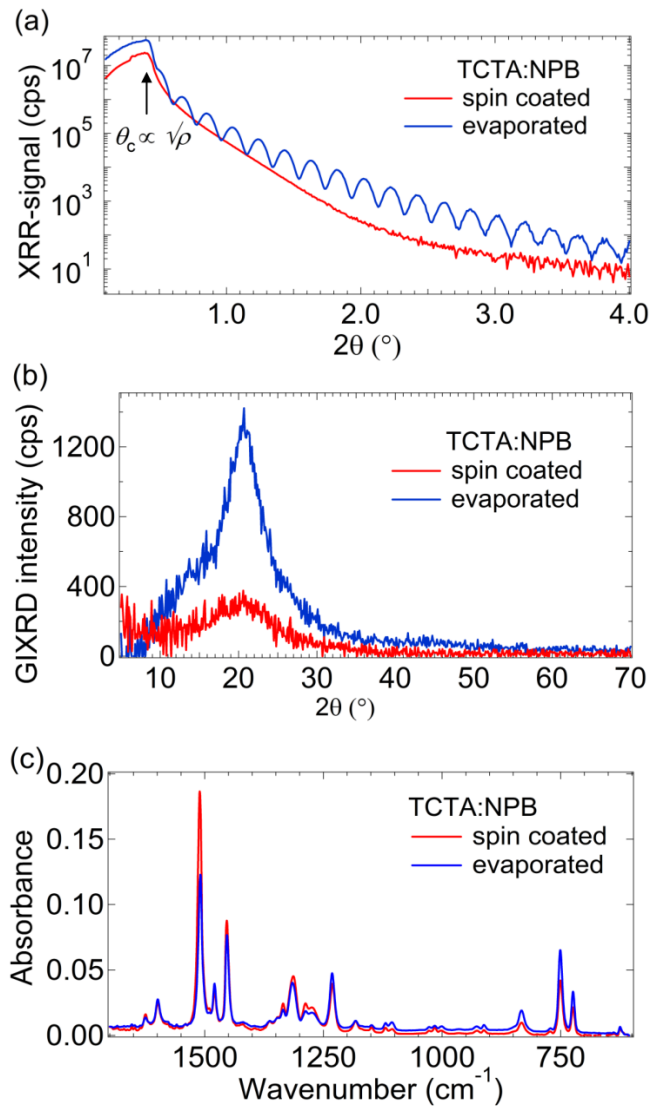

**Figure S2.** Morphology comparison of [TCTA:NPB]$_{sc}$ (red) and [TCTA:NPB]$_{evap}$ (blue) films. (a) X-Ray Reflectivity (XRR) indicating the critical angle $\theta_c$ at the total reflection edge of the XRR signal. (b) Grazing Incidence X-ray Diffraction (GIXRD) demonstrating broad diffraction peaks for both systems at $2\theta \sim 22°$. (c) Fourier transform infrared spectroscopy (FTIR) showing the absorbance in the fingerprint region (1500 to 500 cm$^{-1}$), obtained in Attenuated total reflection (ATR) mode.

According to the Fourier transform infrared spectroscopy (FTIR) spectra of TCTA:NPB films, deposited via spin coating (red) and evaporation (blue) techniques, the vibration modes are not influenced by the processing technique, but it rather presents different magnitudes throughout the IR spectrum (cf. Figure S2c), which indicates different concentration of bonds that contribute to each frequency [3]. We have selected the fingerprint region (from about 1500 to 500 cm$^{-1}$) for comparison, because this is the region of the IR in which a distinct pattern to each different compound is produced [3] and therefore, by observing similar spectrum for both samples (spin coated and thermal evaporated) we are able to prove that the molecules are not affected by the processing technique.

## 3. TTA-based delayed fluorescence

We have observed delayed fluorescence in the delayed PL spectra (cf. Figure S3), which we appoint to annihilation-based up-conversion (triplet-triplet-annihilation) [4-5]. According to the PL spectra, the delayed fluorescence is as high as 50, 33, and 0.5% the phosphorescence of [TCTA:NPB]$_{evap}$ (blue), [TCTA:NPB]$_{sc}$ (red), and PMMA:NPB (black), respectively. We fitted the experimental data of the fluorescence spectra for all systems to the delayed fluorescence (DF) peak. From the observations, we can deduce that in all cases, the delayed fluorescence does not show a spectral shift compared to the fluorescence emission. However, the fluorescence spectrum in TCTA:NPB system showed a slight red shifted emission compared to that of PMMA:NPB system, indicating a presence of interactions between NPB molecule and TCTA matrix. For example, in the case of PMMA:NPB system, the singlet excitons are directly generated by light-absorption, but in the TCTA:NPB system, singlet excitons of NPB are mostly created through a resonance energy transfer from singlet of TCTA to singlet of NPB because excitation light is mainly absorbed by TCTA host molecules, as shown Figure S3.1a and S3.1b. Moreover, the overlap between the emission of TCTA and absorbance of NPB is favorable for a donor (TCTA)-acceptor (NPB) energy transfer case (cf. Figure S3.1c).

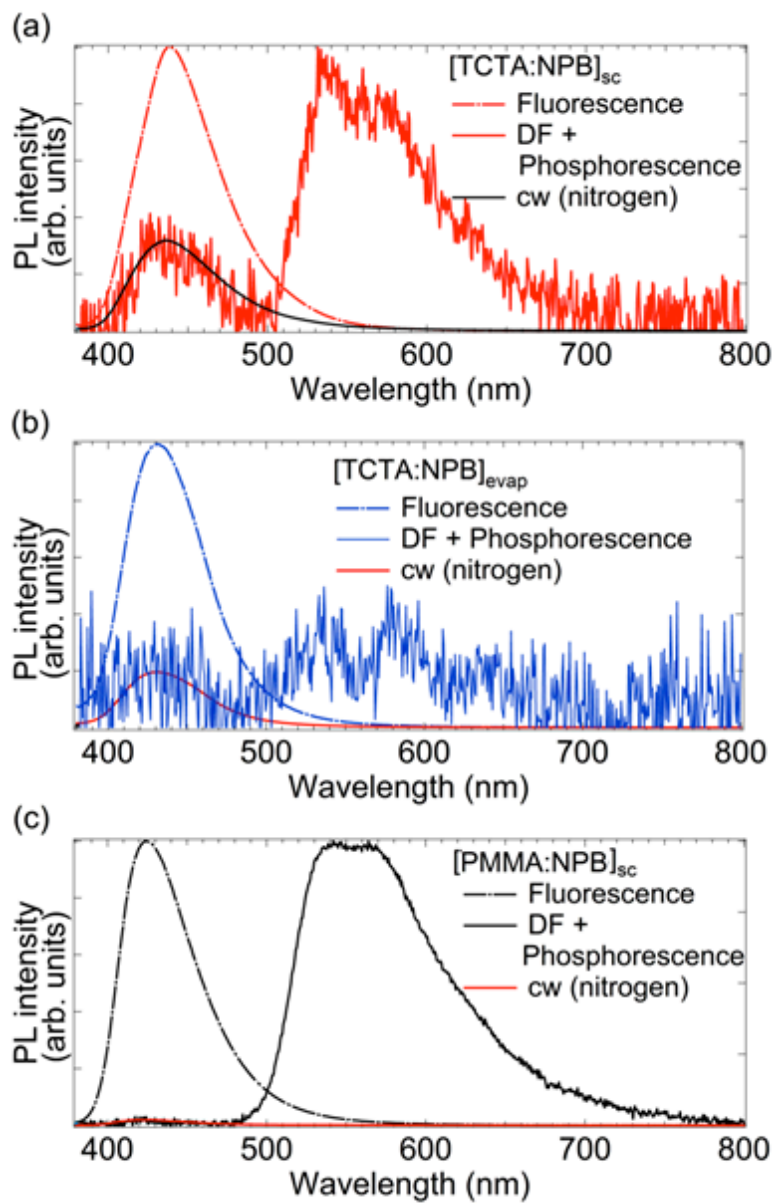

**Figure S3.** Continuous wave (fluorescence, dashed line) and gated (delayed fluorescence "DF" and phosphorescence) photoluminescence spectra under nitrogen atmosphere at room temperature of (a) [TCTA:NPB]$_{sc}$ (red), (b) [TCTA:NPB]$_{evap}$ (blue) and (c) [PMMA:NPB]$_{sc}$ (black). The delayed spectrum of all systems includes their fluorescence spectrum (cw (nitrogen)) with the aim of fitting it to its counterpart delayed fluorescence.

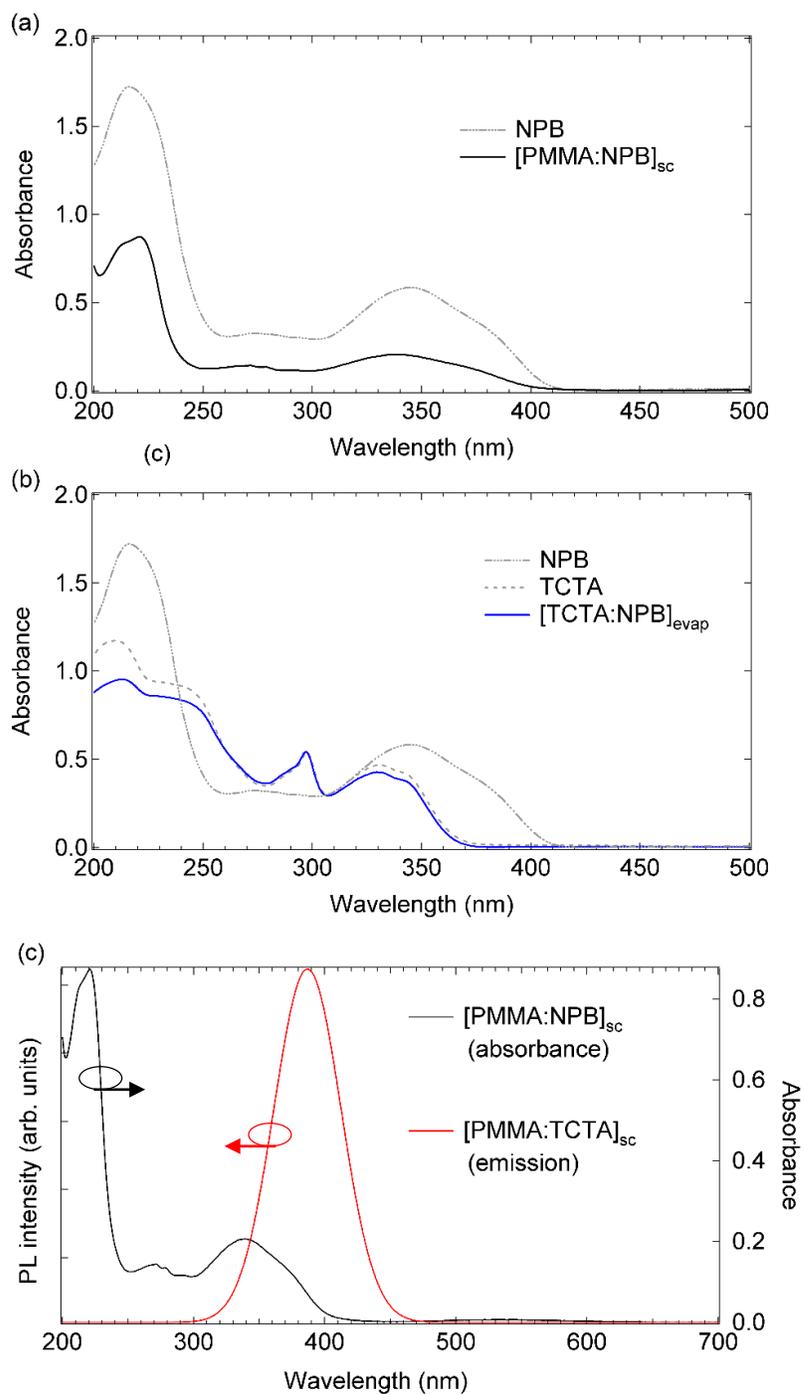

**Figure S3.1** Absorbance spectrum of (a) NPB (grey dotted line), PMMA:NPB (black line), (b) TCTA (grey dashed line), and TCTA:NPB (blue line) thin films. (c) Overlap between absorbance spectrum of NPB (black line) and emission spectrum of TCTA (red line). TCTA fluorescence spectrum, as plotted here, represents a Gaussian fit to the experimental fluorescence spectrum (for clarity). The latter is superimposed by the LED excitation source (365 nm).

# 4. Calculation of the fluorescence and phosphorescence quantum yield

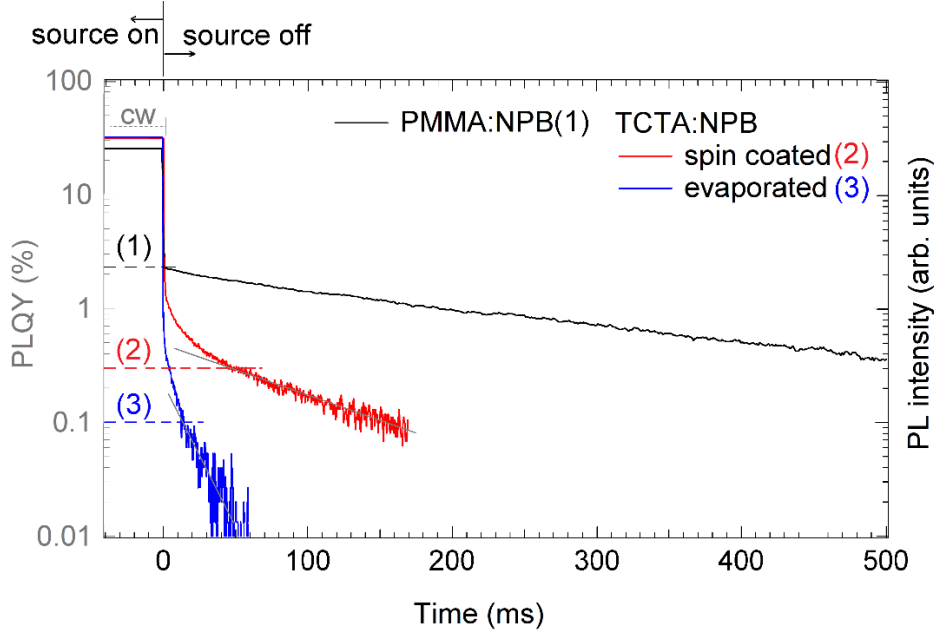

**Figure S4.** Time dependent emission profile of biluminescent films of NPB embedded in PMMA (black), spin coated TCTA (red) and thermal evaporated TCTA (blue). Phosphorescence transients following a pulsed LED with periods of 1 s, 300 ms and 100 ms for [PMMA:NPB]$_{sc}$, [TCTA:NPB]$_{sc}$ and [TCTA:NPB]$_{evap}$ samples, respectively, and 50% duty cycle. Data of the respective samples under cw-illumination (data points when Time < 0) are scaled to the PL quantum yield (PLQY, grey y-axis). Labels (1), (2) and (3) correspond to the phosphorescence quantum yield (PLQY$_P$) of [PMMA:NPB]$_{sc}$, [TCTA:NPB]$_{sc}$ and [TCTA:NPB]$_{evap}$, respectively.

The intensity between $t = -460$ ms and $t = 0$ (source on) corresponds to the total emission of singlet (fluorescence) and triplet (phosphorescence) at steady state under cw-illumination. Since the PLQY is also obtained by integrating the total intensity of fluorescence and phosphorescence under cw-illumination during 1 s, which is enough for the systems to reach steady state, one can assume that the PL intensity at -460 ms <Time < 0 (source on) is related to the PLQY. Therefore, the PL intensity of each system is normalized to their corresponding PLQY in order to estimate the fluorescence (PLQY$_F$) and phosphorescence (PLQY$_P$) yields as follows.

When Time >0 (source off), the long lived phosphorescence is observed. The value at the phosphorescence onset corresponds to the initial phosphorescence intensity which at the same time is related to $PLQY_P$. The phosphorescence decay of [TCTA:NPB]sc and [TCTA:NPB]evap is multi-exponential, due to the TTA-based delayed fluorescence (cf. Figure S3), which influences the shape during the first milliseconds, therefore, the chosen $PLQY_P$ value is extracted out of the contribution from the linear part of the decay.

Finally, the fluorescence yield is the remaining part of the PLQY as follows:

$$PLQY_F = PLQY - PLQY_P \qquad (3)$$

All values are summarized in the following table:

| System | PLQY [%] | $PLQY_F$ | $PLQY_P$ |
|---|---|---|---|
| $[PMMA:NPB]_{sc}$ | 26±3 | 23.7 | 2.3 |
| $[TCTA:NPB]_{sc}$ | 31±8 | 30.7 | 0.3 |
| $[TCTA:NPB]_{evap}$ | 32±3 | 31.9 | 0.1 |

Although, we can obtain a rough estimation of the fluorescence and phosphorescence yield using this technique, it is not completely accurate, because here, we assume that the samples have the same absorption in both measurements (PLQY and time resolved PL), which is not true, although the integration time was the same, the excitation power are not equal in both cases and the phosphorescence onset is arbitrarily chosen. Therefore, to obtain a more precise value, the phosphorescence yield can be calculated as follows. The PLQY represents the total efficiency of the system, in order to calculate the separate phosphorescence and fluorescence efficiencies, one must separate both emissions spectra completely. If the total emission of fluorescence and phosphorescence is defined as the total integrated spectrum at every wavelength (380 < λ < 800 nm) in steady state during cw-illumination ($I_{TOTAL}$), and at the same time, the total emission intensity at every wavelength (380 < λ < 800 nm) after the excitation source is off is defined as the delayed intensity ($I_{DELAYED}$), which corresponds to the phosphorescence ($I_P$) and delayed fluorescence ($I_{DF}$) emissions. Then, the delayed yield can be calculated as:

$$PLQY_D = PLQY \left(\frac{I_{DELAYED}}{I_{TOTAL}}\right) \qquad (4)$$

Additionally, we can assign the contributions of both delayed fluorescence ($A_{DF}$) and phosphorescence ($A_P$) to the fractions of short and long lifetime components (i.e. $A_1$ and $A_2$ of Equation (1)) for each corresponding phosphorescence transient data, respectively. Finally, the phosphorescence yield is computed using Equation (5):

$$PLQY_P = A_P PLQY_D \qquad (5)$$

In this method, one must ensure that the spectra $I_{TOTAL}$ and $I_{DELAYED}$ are acquired with the same integration time and excitation surface power. In this case, those parameters were set to 450 ms and 2.6 x $10^3$ J/s.m², respectively. The excitation surface power was derived from the LED power (360 mW) and the excited area in the sample (⌀ = 25 mm) as a function of the transmission of the bandpass filter used to limit the LED emission (%T = 35).

The values obtained here are similar to the ones estimated from Figure S4, because the excitation surface energy are comparable in both measurements (PLQY and spectral resolved PL) $25.2 \times 10^3$ J/m$^2$ and $5.7 \times 10^3$ J/m$^2$ respectively.

All these values are summarized in the following table:

| System | PLQY [%] | PLQY$_F$ [%] | PLQY$_D$ [%] | PLQY$_P$ [%] | A$_{DF}$ | A$_P$ |
|---|---|---|---|---|---|---|
| [PMMA:NPB]$_{sc}$ | 26±3 | 23.1 | 0.3 | 2.6 | 0.1 | 0.9 |
| [TCTA:NPB]$_{sc}$ | 31±8 | 30.4 | 0.2 | 0.4 | 0.4 | 0.6 |
| [TCTA:NPB]$_{evap}$ | 32±3 | 31.6 | 0.2 | 0.2 | 0.5 | 0.5 |

## 5. Experimental determination of the singlet and triplet population of the biluminescent system PMMA:NPB

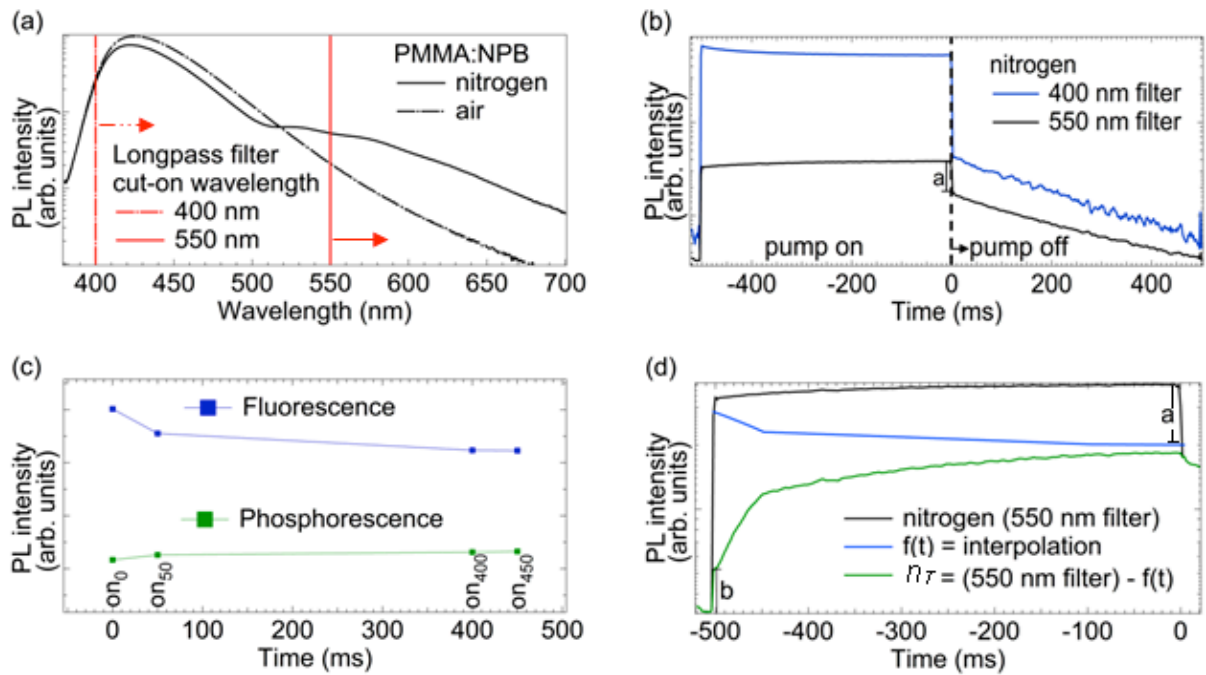

**Figure S5**. Experimental estimation of the singlet and triplet exciton densities of PMMA:NPB. (a) Photoluminescence spectra under continuous wave excitation, in both air (black dashed line) and nitrogen (black solid line). Red solid and dashed lines at 550 and 400 nm respectively, indicate the cut-on wavelengths of the longpass filters used during the time resolved experiments. (b) PL intensity during and after a 500 ms excitation pulse. The intensity was detected through a 400 nm (blue) and 550 nm (black) filters, suggesting the development of singlet and triplet populations during quasi-cw excitation pulses respectively. Constant "a" indicates the drop in fluorescence intensity immediately after the excitation source is off. (c) Intensities of fluorescence (380 – 500 nm) and phosphorescence (500 – 700 nm) obtained in 20 ms integration windows with different delays to the pump pulse start. Here, $on_0$ = no delay, $on_{50}$ = 50 ms delay, $on_{400}$ = 400 ms delay, and $on_{450}$ = 450 ms delay. (d) Final profile of the triplet population evolution over time (green), as result of the subtraction of the interpolated fluorescence (blue) to the experimental data using the 550 nm filter (black). Constant "b" is attributed to the statistical error due to crosstalk between fluorescence and phosphorescence emissions.

We observed a correlation between the evolution of the measured PL intensity over time with the singlet and triplet densities of NPB. We chose the reference system PMMA:NPB to investigate this further.

In order to estimate experimentally the population of the singlet and triplet manifolds, we used time resolved photoluminescence measurements under nitrogen atmosphere. Long pass filters with cut-on wavelengths at 400 and 550 nm were used in front of the photodetector to selectively detect fluorescence and phosphorescence intensities respectively. However, the fluorescence emission extends to 700 nm (dashed black line in Figure S4a), so there is a fraction of fluorescence intensity which passes through the 550 nm filter, seen as the tail at the right side of the red solid line (cf. Figure S4a). This induces a contribution of the fluorescence in the measurement which aims to detect the phosphorescence emission. Such fluorescence contribution is observed in the PL transient measurement with the 550 nm filter (black solid line in Figure S4b) as the drop of intensity at "Time=0" (when excitation pump is switched off), and it is labeled as "a" (cf. Figure S4b).

Figure S4c shows the PL response over time of the PMMA:NPB samples as they were exposed to 500 ms on, 500 ms off cycles quasi-cw pulse. When "pump on" the intensity over time of the fluorescence decreases (blue solid line) and at the same time the phosphorescence increases (black solid line), suggesting a profile of the singlet and triplet population density until a steady state is reached, before the excitation pulse is off. This result was reproduced as obtained from spectrally resolved time-gated measurements (cf. Figure S4c). Here, the intensities of fluorescence (380 – 500 nm) and phosphorescence (500 – 700 nm) were obtained in 20 ms integration windows with different delays to the starting point of the excitation pulse, where, $on_0$ = no delay, $on_{50}$ = 50 ms delay, $on_{400}$ = 400 ms delay, and $on_{450}$ = 450 ms delay.

A profile which is closer to the population density of the triplet state $n_T$ is shown in Figure S4d (green line). Here, the increase in intensity represents the increase of triplets over time while the molecules are pumped, it starts to decrease from "Time=0" once the pump is off. Since there is a contribution of the fluorescence intensity in the phosphorescence measurements, $n_T$ can be estimated if: The fluorescence data points in Figure S4c are interpolated to get a singlet density function "f(t)" (blue line, Figure S4d), which is normalized to the leaked fluorescence intensity "a" and subtracted to the measured phosphorescence (black line, Figure S4d), as expressed in:

$$n_{T\_experimental} \propto I_{phosphorescence} - I_{leaked\_fluorescence} \qquad (3)$$

Where $n_T$ experimental is the experimental population density of the triplet state, $I_{phosphorescence}$ is the measured intensity using the 550 nm longpass filter, and $I_{leaked\_fluorescence}$ is the fraction of fluorescence intensity above 550 nm which influences the measured intensity considered as the phosphorescence emission.

The triplet density related to the phosphorescence intensity does not start at zero. The remaining intensity "b" is attributed to the statistical error of the post-processing, needed due to the overlap between fluorescence and phosphorescence emission of the biluminescent emitter NPB.

**Supporting references**